\documentclass[aps,preprint]{revtex4}%
\usepackage{amsfonts}
\usepackage{amsmath}
\usepackage{amssymb}
\usepackage{graphicx}%
\newtheorem{theorem}{Theorem}
\newtheorem{acknowledgement}[theorem]{Acknowledgement}

\newtheorem{proposition}[theorem]{Proposition}

\newenvironment{proof}[1][Proof]{\noindent\textbf{#1.} }{\ \rule{0.5em}{0.5em}}
\begin{document}
\preprint{ }
\title[ ]{No-Go Theorem for Energy-Momentum Conservation in Curved Spacetime}
\author{Zhaoyan Wu}
\email{zhaoyanwu2000@yahoo.com}
\affiliation{Center for Theoretical Physics, Jilin University, China}

\begin{abstract}
It is pointed out that in curved spacetime, one cannot define the sum of
energy-momemtum 4-vector over a space-like hypersurface. The difficulty in
finding satisfactory gravitational energy-momentum complex stems from
misunderstanding of this question. The law of conservation of energy-momentum
holds valid only approximately when spacetime is not seriously curved.

\end{abstract}
\maketitle

\section{\label{intro}Introduction}

The law of conservation of energy is the cornerstone of physics. As a matter
of fact, far beyond physics, it governs all natural processes, and is
considered the bedrock law of nature. In 1686, Leibniz proposed the concept of
kinetic energy, when he noticed that the total kinetic energy remains
unchanged in some processes, say, elastic collissions on a horizontal
frictionless plane. Later, however, it was found that this was not true for
some other processes. The concept of\ potential energy (elastic,
gravitational, etc.) was then introduced, so that the total mechanic energy
might remain unchanged. To keep the total energy unchanged, energy of heat,
electromagnetic energy, chemical energy, etc., were introduced. People
realized that energy can not be created nor be destroyed, it can only change
its forms, and be transported from body to body. In a sense, the processes of
development of physics since, can be regarded as the processes of discovery of
new forms and new carriers of energy in order to keep the law of energy
conservation alive. The discovery of neutino in 1930's to keep conservation of
energy-momentum valid in $\beta$-decay, was a good example.

In 1905, more than two centuries after Leibniz proposed the concept of kinetic
energy, Einstein presented his special theory of relativity, which changed
revolutionarily the concepts of space and time. Space and time can no longer
be separated absolutely. And energy is no longer a scalar, but the
$0$-component of the energy-momentum $4$-vector. Conservation of energy and
conservation of momentum either both hold valid, or both fail, because
conservation of energy (momentum) is the consequence of homogeneity of time
(space), and spacetime is inseparable. In special relativity (SR) spacetime is
still flat, and we have conservation of energy-momentum:%

\begin{equation}
\partial_{\mu}T^{\mu\nu}(x)=0 \tag{1}%
\end{equation}%
\begin{equation}
\int_{\Omega}d^{4}x\partial_{\mu}T^{\mu\nu}(x)=\int_{\partial\Omega}ds_{\mu
}T^{\mu\nu}(x)=0 \tag{2}%
\end{equation}
where $\Omega$ is a 4-dimensional spacetime region, $\partial\Omega$ its
boundary and $(x^{0},x^{1},x^{2},x^{3})$ a Lorentzian coordinate system
(coordinate system in which the metric tensor $g_{\alpha\beta}(x)\equiv
\eta_{\alpha\beta},$ $[\eta]=diag(-1,1,1,1)$).\ Special relativity also
revealed that mass and energy are the same thing, $mc^{2}=E$.

Ten years later, Einstein published his general theory of relativity.
According to general relativity (GR), our spacetime is no longer flat, but
curved. It is no longer an affine space, but a 4-dimensional generalized
Riemannian manifold with a Lorentzian signatured metric. The law of
energy-momentum conservation encounters difficulty in GR. In fact, from
$\nabla_{\mu}j^{\mu}(x)=0$, which is the generalization to curved spacetime of
$\partial_{\mu}j^{\mu}(x)=0$, the differential conservation law of some scalar
in flat spacetime, we can get the integral conservation law of this scalar in
curved spacetime:%

\begin{equation}
\sqrt{-g(x)}\nabla_{\mu}j^{\mu}(x)=\partial_{\mu}(\sqrt{-g(x)}j^{\mu
}(x))=0\Longrightarrow\int_{\partial\Omega}ds_{\mu}\sqrt{-g(x)}j^{\mu}(x)=0
\tag{3}%
\end{equation}
But, from $\nabla_{\mu}T^{\mu\nu}(x)=0$, the generalization of (1) to curved
spacetime, we cannot get the integral conservation law of energy-momentum in
curved spacetime. In fact,%

\begin{align}
\sqrt{-g(x)}\nabla_{\mu}T^{\mu\nu}(x)  &  =\partial_{\mu}(\sqrt{-g(x)}%
T^{\mu\nu}(x))+\sqrt{-g(x)}\Gamma_{\mu\lambda}^{\nu}T^{\mu\lambda
}=0\nonumber\\
&  \Longrightarrow\int_{\partial\Omega}ds_{\mu}\sqrt{-g(x)}T^{\mu\nu}%
(x)=-\int_{\Omega}d^{4}x\sqrt{-g(x)}\Gamma_{\mu\lambda}^{\nu}T^{\mu\lambda}(x)
\tag{4}%
\end{align}
The rhs of eqn.(4) is not always equal to $0$. Having realized this
difficulty, Einstein rewrote $\nabla_{\mu}T^{\mu\nu}(x)=0$ as[1],%

\begin{equation}
\sqrt{-g(x)}\nabla_{\mu}T^{\mu\nu}(x)=\partial_{\mu}[\sqrt{-g(x)}(T^{\mu\nu
}+t^{\mu\nu})]=0 \tag{5}%
\end{equation}
and by integrating it over $\Omega$, he obtained%

\begin{equation}
\int_{\Omega}d^{4}x\partial_{\mu}[\sqrt{-g(x)}(T^{\mu\nu}+t^{\mu\nu}%
)]=\int_{\partial\Omega}ds_{\mu}[\sqrt{-g(x)}(T^{\mu\nu}+t^{\mu\nu})]=0
\tag{6}%
\end{equation}
which he considered as the integral conservation law of the total
energy-momentum with $t^{\mu\nu}$ being regarded as the energy-momentum tensor
of gravity. However, $t^{\mu\nu}$ does not behave like a tensor and is not
symmetric as expected. H. Bauer pointed out it leads to difficulty[2].
Following Einstein, Tolman[3], Landau, Lifshitz[4], Papapetrou[5],
M\"{o}ller[6], Weinberg[7], Bergmann and Thompson[8] proposed their
gravitational energy-momentum complexes, coordinate dependent or coordinate
free, symmetrical or asymmetrical, but all lacking in energy locality. Bondi
argued that nonlocalizable energy is not allowed in GR[9]. Penrose and many
others developed the concept of quasilocal energy[10]. So far, however, we
still lack a generally accepted definition of energy-momentum of gravitational
field. Energy-momentum conservation in GR and energy-momentum density of
gravity are still the focus of theoretical physicists[11].

The present paper consists of two parts. In the first part, by using
variational principle approach, the field equation in GR is re-established,
and Noether's theorem is re-derived in a context more general than before, and
the conservation laws of energy-momentun, angular momentun and electric charge
are re-obtained. In the second part, a no-go theorem for energy-momentun
conservation in curved spacetime is given. And finally, the physical
implication is discussed.

\section{Variational Priciple Approach to General Relativity}

In the present paper, we use Hilbert action for the metric field, and the
total action is%

\begin{equation}
A=\int_{\Omega}d^{4}x\sqrt{-g(x)}[\frac{1}{16\pi G}R+\mathcal{L}%
_{M}(g(x),u(x),\nabla u(x))]=A_{G}+A_{M} \tag{7}%
\end{equation}
where $R$ is the Ricci scalar curvature, and $\mathcal{L}_{M}$ the Lagrangian
of matter field, which we restrict to scalar and vector fields without loss of
generality for our purpose.

For the sake of generality , we use the following general action in our discussion.%

\begin{equation}
A=\int_{\Omega}d^{4}xL(\Phi(x),\partial\Phi(x),\partial^{2}\Phi(x)) \tag{8}%
\end{equation}

\subsection{Re-establishing Field Equation}

The least action principle says for any 4-dimensional spacetime region
$\Omega$, the action on $\Omega$ of real movement takes the stationary value
among the actions on $\Omega$ of all possible movements (all movements allowed
by the constraints) with the same boundary condition:%

\begin{equation}
\delta\Phi_{a}(x)|_{\partial\Omega}=0 \tag{9}%
\end{equation}
for $\Phi_{a}(x)$'s whose second derivatives$\ \partial^{2}\Phi_{a}(x)$ do not
appear in $L$.%

\begin{equation}
\delta\Phi_{a}(x)|_{\partial\Omega}=0,\delta\partial_{\mu}\Phi_{a}%
(x)|_{\partial\Omega}=0 \tag{10}%
\end{equation}
for $\Phi_{a}(x)$'s whose second derivatives$\ \partial^{2}\Phi_{a}(x)$ do
appear in $L$.

Using the standard procedure, we obtain%

\begin{align}
\delta A  &  =\int_{\Omega}d^{4}x(\frac{\partial L}{\partial\Phi_{a}%
(x)}-\partial_{\mu}\frac{\partial L}{\partial\partial_{\mu}\Phi_{a}%
(x)}+\partial_{\mu}\partial_{\nu}\frac{\partial L}{\partial\partial\mu
\partial_{\nu}\Phi_{a}(x)})\delta\Phi_{a}(x)\nonumber\\
&  +\int_{\Omega}d^{4}x\partial_{\mu}[(\frac{\partial L}{\partial\partial
_{\mu}\Phi_{a}(x)}-\partial_{\nu}\frac{\partial L}{\partial\partial\mu
\partial_{\nu}\Phi_{a}(x)})\delta\Phi_{a}(x)+\frac{\partial L}{\partial
\partial\mu\partial_{\nu}\Phi_{a}(x)}\delta\partial_{\nu}\Phi_{a}%
(x)]\nonumber\\
&  =\int_{\Omega}d^{4}x(\frac{\partial L}{\partial\Phi_{a}(x)}-\partial_{\mu
}\frac{\partial L}{\partial\partial_{\mu}\Phi_{a}(x)}+\partial_{\mu}%
\partial_{\nu}\frac{\partial L}{\partial\partial\mu\partial_{\nu}\Phi_{a}%
(x)})\delta\Phi_{a}(x)\nonumber\\
&  +\int_{\partial\Omega}ds_{\mu}[(\frac{\partial L}{\partial\partial_{\mu
}\Phi_{a}(x)}-\partial_{\nu}\frac{\partial L}{\partial\partial\mu\partial
_{\nu}\Phi_{a}(x)})\delta\Phi_{a}(x)+\frac{\partial L}{\partial\partial
\mu\partial_{\nu}\Phi_{a}(x)}\delta\partial_{\nu}\Phi_{a}(x)]\nonumber\\
&  =\int_{\Omega}d^{4}x(\frac{\partial L}{\partial\Phi_{a}(x)}-\partial_{\mu
}\frac{\partial L}{\partial\partial_{\mu}\Phi_{a}(x)}+\partial_{\mu}%
\partial_{\nu}\frac{\partial L}{\partial\partial\mu\partial_{\nu}\Phi_{a}%
(x)})\delta\Phi_{a}(x)=0 \tag{11}%
\end{align}
Therefore the general field equation is%

\begin{equation}
\frac{\delta A}{\delta\Phi_{a}(x)}=\frac{\partial L}{\partial\Phi_{a}%
(x)}-\partial_{\mu}\frac{\partial L}{\partial\partial_{\mu}\Phi_{a}%
(x)}+\partial_{\mu}\partial_{\nu}\frac{\partial L}{\partial\partial\mu
\partial_{\nu}\Phi_{a}(x)}=0 \tag{12}%
\end{equation}
For the case of general relativity%

\begin{equation}
L=\sqrt{-g(x)}[\frac{1}{16\pi G}R+\mathcal{L}_{M}(g(x),u(x),\nabla
u(x))]=L_{G}+L_{M} \tag{13}%
\end{equation}
we have%

\begin{equation}
\frac{\delta A}{\delta\Phi_{a}(x)}=\frac{\delta A_{G}}{\delta\Phi_{a}%
(x)}+\frac{\delta A_{M}}{\delta\Phi_{a}(x)}=0 \tag{14}%
\end{equation}
When $u(x)$ in (13) is a scalar field,%

\[
\frac{\partial L}{\partial\varphi(x)}=\sqrt{-g(x)}\frac{\partial
\mathcal{L}_{M}}{\partial\varphi(x)}%
\]%
\[
\partial_{\mu}\frac{\partial L}{\partial\partial_{\mu}\varphi(x)}%
=\partial_{\mu}[\sqrt{-g(x)}\frac{\partial\mathcal{L}_{M}}{\partial\nabla
_{\mu}\varphi(x)}]=\sqrt{-g(x)}[\partial_{\mu}\frac{\partial\mathcal{L}_{M}%
}{\partial\nabla_{\mu}\varphi(x)}+\Gamma_{\mu\nu}^{\mu}\frac{\partial
\mathcal{L}_{M}}{\partial\nabla_{\nu}\varphi(x)}]
\]
we get%

\begin{equation}
\frac{\delta A}{\delta\varphi(x)}=\sqrt{-g(x)}[\frac{\partial\mathcal{L}_{M}%
}{\partial\varphi(x)}-\nabla_{\mu}\frac{\partial\mathcal{L}_{M}}%
{\partial\nabla_{\mu}\varphi(x)}]=0 \tag{15}%
\end{equation}
When $u(x)$ in (13) is a vector field,%

\[
\frac{\partial L}{\partial u_{\alpha}(x)}=\sqrt{-g(x)}[\frac{\partial
\mathcal{L}_{M}}{\partial u_{\alpha}(x)}+\frac{\partial\mathcal{L}_{M}%
}{\partial\nabla_{\mu}u_{\beta}(x)}(-\Gamma_{\mu\beta}^{\alpha})]
\]%
\[
\partial_{\mu}\frac{\partial L}{\partial\partial_{\mu}u_{\alpha}(x)}%
=\partial_{\mu}[\sqrt{-g(x)}\frac{\partial\mathcal{L}_{M}}{\partial\nabla
_{\mu}u_{\alpha}(x)}]=\sqrt{-g(x)}[\partial_{\mu}\frac{\partial\mathcal{L}%
_{M}}{\partial\nabla_{\mu}u_{\alpha}(x)}+\Gamma_{\mu\nu}^{\mu}\frac
{\partial\mathcal{L}_{M}}{\partial\nabla_{\nu}u_{\alpha}(x)}]
\]
we get%

\begin{equation}
\frac{\delta A}{\delta u_{\alpha}(x)}=\sqrt{-g(x)}[\frac{\partial
\mathcal{L}_{M}}{\partial u_{\alpha}(x)}-\nabla_{\mu}\frac{\partial
\mathcal{L}_{M}}{\partial\nabla_{\mu}u_{\alpha}(x)}]=0 \tag{16}%
\end{equation}
For both cases, we have%

\begin{equation}
\frac{\delta A}{\delta g^{\alpha\beta}(x)}=\frac{\delta A_{G}}{\delta
g^{\alpha\beta}(x)}+\frac{\delta A_{M}}{\delta g^{\alpha\beta}(x)}=0 \tag{17}%
\end{equation}%
\[
\frac{\partial L_{G}}{\partial g^{\alpha\beta}(x)}=\sqrt{-g(x)}\frac{1}{16\pi
G}[-\frac{1}{2}g_{\alpha\beta}R+\frac{\partial R}{\partial g^{\alpha\beta}%
(x)}]
\]%
\begin{align*}
\partial_{\mu}\frac{\partial L_{G}}{\partial\partial_{\mu}g^{\alpha\beta}(x)}
&  =\partial_{\mu}[\sqrt{-g(x)}\frac{1}{16\pi G}\frac{\partial R}%
{\partial\partial_{\mu}g^{\alpha\beta}(x)}]\\
&  =\sqrt{-g(x)}\frac{1}{16\pi G}[\partial_{\mu}\frac{\partial R}%
{\partial\partial_{\mu}g^{\alpha\beta}(x)}-\frac{1}{2}g_{\rho\sigma}%
\partial_{\mu}g^{\rho\sigma}\frac{\partial R}{\partial\partial_{\mu}%
g^{\alpha\beta}(x)}]
\end{align*}%
\begin{align*}
\partial_{\mu}\partial_{\nu}\frac{\partial L_{G}}{\partial\partial_{\mu
}\partial_{\nu}g^{\alpha\beta}(x)}  &  =\partial_{\mu}\partial_{\nu}%
[\sqrt{-g(x)}\frac{1}{16\pi G}\frac{\partial R}{\partial\partial_{\mu}%
\partial_{\nu}g^{\alpha\beta}(x)}]\\
&  =\sqrt{-g(x)}\frac{1}{16\pi G}[\frac{1}{4}g_{\rho\sigma}\partial_{\mu
}g^{\rho\sigma}g_{\xi\eta}\partial_{\nu}g^{\xi\eta}\frac{\partial R}%
{\partial\partial_{\mu}\partial_{\nu}g^{\alpha\beta}(x)}-g_{\rho\sigma
}\partial_{\mu}g^{\rho\sigma}\partial_{\nu}\frac{\partial R}{\partial
\partial_{\mu}\partial_{\nu}g^{\alpha\beta}(x)}\\
&  -\frac{1}{2}g_{\rho\sigma}\partial_{\mu}\partial_{\nu}g^{\rho\sigma}%
\frac{\partial R}{\partial\partial_{\mu}\partial_{\nu}g^{\alpha\beta}%
(x)}+\frac{1}{2}g_{_{\rho\xi}}g_{\sigma\eta}\partial_{\mu}g^{\rho\sigma
}\partial_{\nu}g^{\xi\eta}\frac{\partial R}{\partial\partial_{\mu}%
\partial_{\nu}g^{\alpha\beta}(x)}+\partial_{\mu}\partial\nu\frac{\partial
R}{\partial\partial_{\mu}\partial_{\nu}g^{\alpha\beta}(x)}]
\end{align*}
Substituting these equations into eqn.(17), we get%

\begin{equation}
\frac{\delta A}{\delta g^{\alpha\beta}(x)}=\sqrt{-g(x)}\frac{1}{16\pi
G}[R_{\alpha\beta}-\frac{1}{2}g_{\alpha\beta}R-8\pi GT_{\alpha\beta}+I]=0
\tag{18}%
\end{equation}
where%

\begin{equation}
T_{\alpha\beta}\equiv\frac{-2}{\sqrt{-g(x)}}\frac{\delta A_{M}}{\delta
g^{\alpha\beta}(x)} \tag{19}%
\end{equation}
is the energy-momemtum tensor of the matter field, and%

\begin{align*}
I  &  \equiv-R_{\alpha\beta}+\frac{\partial R}{\partial g^{\alpha\beta}%
(x)}-\partial_{\mu}\frac{\partial R}{\partial\partial_{\mu}g^{\alpha\beta}%
(x)}+\partial_{\mu}\partial\nu\frac{\partial R}{\partial\partial_{\mu}%
\partial_{\nu}g^{\alpha\beta}(x)}+\frac{1}{2}g_{\rho\sigma}\partial_{\mu
}g^{\rho\sigma}\frac{\partial R}{\partial\partial_{\mu}g^{\alpha\beta}(x)}\\
&  +\frac{1}{4}g_{\rho\sigma}\partial_{\mu}g^{\rho\sigma}g_{\xi\eta}%
\partial_{\nu}g^{\xi\eta}\frac{\partial R}{\partial\partial_{\mu}\partial
_{\nu}g^{\alpha\beta}(x)}-g_{\rho\sigma}\partial_{\mu}g^{\rho\sigma}%
\partial_{\nu}\frac{\partial R}{\partial\partial_{\mu}\partial_{\nu}%
g^{\alpha\beta}(x)}-\frac{1}{2}g_{\rho\sigma}\partial_{\mu}\partial_{\nu
}g^{\rho\sigma}\frac{\partial R}{\partial\partial_{\mu}\partial_{\nu}%
g^{\alpha\beta}(x)}%
\end{align*}%
\begin{equation}
+\frac{1}{2}g_{_{\rho\xi}}g_{\sigma\eta}\partial_{\mu}g^{\rho\sigma}%
\partial_{\nu}g^{\xi\eta}\frac{\partial R}{\partial\partial_{\mu}\partial
_{\nu}g^{\alpha\beta}(x)}\equiv0 \tag{20}%
\end{equation}
Identity (20) can be proved straightforwardly. Therefore, eqn.(18) is just
Einstein's field equation.%

\begin{equation}
\frac{\delta A}{\delta g^{\alpha\beta}(x)}=\sqrt{-g(x)}\frac{1}{16\pi
G}[(R_{\alpha\beta}-\frac{1}{2}g_{_{\alpha\beta}}R)-8\pi GT_{\alpha\beta}]=0
\tag{21}%
\end{equation}
Thus, eqns.(15)$+$(21) (eqns.(16)$+$(21)) are the field equations of scalar
(vector) field\ plus metric\ field.

\subsection{Re-deriving Noether's Theorem}

Noether's theorem is re-derived in a more general context than before. And by
using it, we get the conservation laws of `energy-momentum' `angular momentum'
and electric charge in general relativity.

\begin{theorem}
Noether's theorem
\end{theorem}

\textit{If the action of field system on any }$4-$dimensional spacetime region
$\Omega$ \textit{remains unchanged under the }$r-$parameter family of
\textit{the following infinitesimal transformations of coordinates and fields}%

\begin{equation}
x^{\mu}\longmapsto\widetilde{x}^{\mu}=x^{\mu}+\delta x^{\mu} \tag{22}%
\end{equation}%
\begin{equation}
\Phi_{a}(x)\mapsto\widetilde{\Phi_{a}}(\widetilde{x})=\Phi_{a}(x)+\delta
\Phi_{a}(x) \tag{23}%
\end{equation}
\textit{then there exist }$r$\textit{ conserved quantities.}

\begin{proof}
From equations (22) and (23), we have%

\begin{equation}
\delta(d^{4}x)=(\partial_{\sigma}\delta x^{\sigma})d^{4}x \tag{24}%
\end{equation}%
\[
\delta\partial_{\mu}=-(\partial_{\mu}\delta x^{\sigma})\partial_{\sigma}%
\]%
\begin{equation}
\delta\partial_{\mu}\Phi_{a}(x)=\partial_{\mu}\delta\Phi_{a}(x)-\partial
_{\sigma}\Phi_{a}(x)(\partial_{\mu}\delta x^{\sigma}) \tag{25}%
\end{equation}%
\begin{equation}
\delta\partial_{\mu}\partial_{\nu}\Phi_{a}(x)=\partial_{\mu}\partial_{\nu
}\delta\Phi_{a}(x)-\partial_{\sigma}\partial_{\nu}\Phi_{a}(x)\partial_{\mu
}\delta x^{\sigma}-\partial_{\sigma}\partial_{\mu}\Phi_{a}(x)\partial_{\nu
}\delta x^{\sigma}-\partial_{\sigma}\Phi_{a}(x)\partial_{\mu}\partial_{\nu
}\delta x^{\sigma} \tag{26}%
\end{equation}
The variation of action is%

\begin{align*}
\delta A  &  =\int_{\Omega}d^{4}x[\frac{\partial L}{\partial\Phi_{a}(x)}%
\delta\Phi_{a}(x)+\frac{\partial L}{\partial\partial_{\mu}\Phi_{a}%
(x)}(\partial_{\mu}\delta\Phi_{a}(x)-\partial_{\sigma}\Phi_{a}(x)\partial
_{\mu}\delta x^{\sigma})+\\
&  +\frac{\partial L}{\partial\partial_{\mu}\partial_{\nu}\Phi_{a}%
(x)}(\partial_{\mu}\partial_{\nu}\delta\Phi_{a}(x)-\partial_{\sigma}%
\partial_{\nu}\Phi_{a}(x)\partial_{\mu}\delta x^{\sigma}-\partial_{\sigma
}\partial_{\mu}\Phi_{a}(x)\partial_{\nu}\delta x^{\sigma}\\
&  -\partial_{\sigma}\Phi_{a}(x)\partial_{\mu}\partial_{\nu}\delta x^{\sigma
})+L\partial_{\sigma}\delta x^{\sigma}]\\
&  =\int_{\Omega}d^{4}x[(\frac{\partial L}{\partial\Phi_{a}(x)}-\partial_{\mu
}\frac{\partial L}{\partial\partial_{\mu}\Phi_{a}(x)}+\partial_{\mu}%
\partial_{\nu}\frac{\partial L}{\partial\partial_{\mu}\partial_{\nu}\Phi
_{a}(x)})(\delta\Phi_{a}(x)-\partial_{\sigma}\Phi_{a}(x)\delta x^{\sigma})]\\
&  +\int_{\Omega}d^{4}x\partial_{\mu}[\delta_{\sigma}^{\mu}L\delta x^{\sigma
}+(\frac{\partial L}{\partial\partial_{\mu}\Phi_{a}(x)}-\partial_{\nu}%
\frac{\partial L}{\partial\partial_{\mu}\partial_{\nu}\Phi_{a}(x)})(\delta
\Phi_{a}(x)-\partial_{\sigma}\Phi_{a}(x)\delta x^{\sigma})
\end{align*}%
\begin{equation}
+\frac{\partial L}{\partial\partial_{\mu}\partial_{\nu}\Phi_{a}(x)}%
\partial_{\nu}(\delta\Phi_{a}(x)-\partial_{\sigma}\Phi_{a}(x)\delta x^{\sigma
})]=0 \tag{27}%
\end{equation}
The first integral at rhs vanishes for real movement, hence the second
integral at rhs does too. We get the following equation due to the
arbitrariness of $\Omega$.%

\begin{align}
&  \partial_{\mu}[\delta_{\sigma}^{\mu}L\delta x^{\sigma}+(\frac{\partial
L}{\partial\partial_{\mu}\Phi_{a}(x)}-\partial_{\nu}\frac{\partial L}%
{\partial\partial_{\mu}\partial_{\nu}\Phi_{a}(x)})(\delta\Phi_{a}%
(x)-\partial_{\sigma}\Phi_{a}(x)\delta x^{\sigma})\nonumber\\
&  +\frac{\partial L}{\partial\partial_{\mu}\partial_{\nu}\Phi_{a}(x)}%
\partial_{\nu}(\delta\Phi_{a}(x)-\partial_{\sigma}\Phi_{a}(x)\delta x^{\sigma
})]\nonumber\\
&  =0 \tag{28}%
\end{align}
Noting that both $\delta\Phi_{a}(x)$ and $\delta x^{\mu}$ depend on $r$ real
parameters, we can consider equation (28) as the conservation laws of $r$ quantities.
\end{proof}

\subsection{Conservation of `Energy-Momentum'}

Action (7) remains unchanged under the following coordinates shift.%

\begin{equation}
x^{\mu}\mapsto\widetilde{x}^{\mu}=x^{\mu}+\varepsilon^{\mu},\ \ \delta x^{\mu
}=\varepsilon^{\mu} \tag{29}%
\end{equation}%
\begin{equation}
\delta u_{\alpha}(x)=0,\ \ \delta g^{\alpha\beta}(x)=0 \tag{30}%
\end{equation}
In this case, eqn.(28) reads%

\begin{equation}
\partial_{\mu}[\delta_{\sigma}^{\mu}L-(\frac{\partial L}{\partial\partial
_{\mu}\Phi_{a}(x)}-\partial_{\nu}\frac{\partial L}{\partial\partial_{\mu
}\partial_{\nu}\Phi_{a}(x)})\partial_{\sigma}\Phi_{a}(x)-\frac{\partial
L}{\partial\partial_{\mu}\partial_{\nu}\Phi_{a}(x)}\partial_{\nu}%
\partial_{\sigma}\Phi_{a}(x)]=0 \tag{31}%
\end{equation}
In order to find out the gravitational energy-momentum complex, for the sake
of simplicity, we consider the complex scalar field in curved spacetime%

\begin{equation}
\mathcal{L}_{M}=-g^{\rho\sigma}\nabla_{\rho}\varphi^{\ast}\nabla_{\sigma
}\varphi-m^{2}\varphi^{\ast}\varphi\tag{32}%
\end{equation}
In this case, eqn.(31) reads%

\begin{align}
&  \partial_{\mu}\{\sqrt{-g(x)}[\delta_{\nu}^{\mu}\mathcal{L}_{M}%
-\frac{\partial\mathcal{L}_{M}}{\partial\nabla_{\mu}\varphi(x)}\nabla_{\nu
}\varphi(x)-\frac{\partial\mathcal{L}_{M}}{\partial\nabla_{\mu}\varphi^{\ast
}(x)}\nabla_{\nu}\varphi^{\ast}(x)]+\sqrt{-g(x)}\frac{1}{16\pi G}[\delta_{\nu
}^{\mu}R\nonumber\\
&  -(\frac{\partial R}{\partial\partial_{\mu}g^{\alpha\beta}(x)}%
-\partial_{\sigma}\frac{\partial R}{\partial\partial_{\mu}\partial_{\sigma
}g^{\alpha\beta}(x)}+\frac{1}{2}g_{\xi\eta}(x)\partial_{\sigma}g^{\xi\eta
}(x)\frac{\partial R}{\partial\partial_{\mu}\partial_{\sigma}g^{\alpha\beta
}(x)})\partial_{\nu}g^{\alpha\beta}(x)\nonumber\\
&  -\frac{\partial R}{\partial\partial_{\mu}\partial_{\sigma}g^{\alpha\beta
}(x)}\partial_{\nu}\partial_{\sigma}g^{\alpha\beta}(x)]\}\nonumber\\
&  =0,\ \ \forall\nu=0,1,2,3 \tag{33}%
\end{align}
The expression in the first bracket is just the enegy-momentum tensor of
matter fields.%

\begin{equation}
T_{\nu}^{\mu}=\delta_{\nu}^{\mu}\mathcal{L}_{M}-\frac{\partial\mathcal{L}_{M}%
}{\partial\nabla_{\mu}\varphi(x)}\nabla_{\nu}\varphi(x)-\frac{\partial
\mathcal{L}_{M}}{\partial\nabla_{\mu}\varphi^{\ast}(x)}\nabla_{\nu}%
\varphi^{\ast}(x) \tag{34}%
\end{equation}
Hence the second bracket with its coefficient%

\begin{align}
t_{\nu}^{\mu}  &  =\frac{1}{16\pi G}[\delta_{\nu}^{\mu}R-(\frac{\partial
R}{\partial\partial_{\mu}g^{\alpha\beta}(x)}-\partial_{\sigma}\frac{\partial
R}{\partial\partial_{\mu}\partial_{\sigma}g^{\alpha\beta}(x)}\nonumber\\
&  +\frac{1}{2}g_{\xi\eta}(x)\partial_{\sigma}g^{\xi\eta}(x)\frac{\partial
R}{\partial\partial_{\mu}\partial_{\sigma}g^{\alpha\beta}(x)})\partial_{\nu
}g^{\alpha\beta}(x)-\frac{\partial R}{\partial\partial_{\mu}\partial_{\sigma
}g^{\alpha\beta}(x)}\partial_{\nu}\partial_{\sigma}g^{\alpha\beta}(x)]
\tag{35}%
\end{align}
seems to be the gravitational counterpart. It is not difficult to show that
eqn.(34) is consistent with eqn.(19) for Lagrangian (32).

\subsection{Conservation of `Angular Momentum'}

Action (7) remains unchanged under the following infinitesimal `Lorentz
transformation' (36) through (38). In this subsection, $u(x)$ in (7) is
supposed to be the complex scalar field for definiteness.%

\[
x^{\mu}\mapsto\widetilde{x}^{\mu}=L_{\nu}^{\mu}x^{\nu},\ L_{\nu}^{\mu}%
=\delta_{\nu}^{\mu}+\Lambda_{\nu}^{\mu},\ |\Lambda_{\nu}^{\mu}|\ll1,\eta
_{\mu\lambda}\Lambda_{\nu}^{\lambda}\equiv\Lambda_{\mu\nu},\Lambda_{\mu
\lambda}=-\Lambda_{\lambda\mu}%
\]%
\begin{equation}
\delta x^{\mu}=\Lambda_{\nu}^{\mu}x^{\nu}=\frac{1}{2}(\eta^{\mu\rho}x^{\sigma
}-\eta^{\mu\sigma}x^{\rho})\Lambda_{\rho\sigma} \tag{36}%
\end{equation}%
\begin{equation}
\delta\varphi(x)=0,\ \delta\varphi^{\ast}(x)=0 \tag{37}%
\end{equation}%
\begin{align}
\delta g^{\alpha\beta}(x)  &  =\Lambda_{\sigma}^{\beta}g^{\alpha\sigma
}+\Lambda_{\sigma}^{\alpha}g^{\sigma\beta}=\eta^{\beta\rho}\Lambda_{\rho
\sigma}g^{\alpha\sigma}+\eta^{\alpha\rho}\Lambda_{\rho\sigma}g^{\sigma\beta
}\nonumber\\
&  =\frac{1}{2}(\eta^{\beta\rho}g^{\alpha\sigma}-\eta^{\beta\sigma}%
g^{\alpha\rho}+\eta^{\alpha\rho}g^{\sigma\beta}-\eta^{\alpha\sigma}%
g^{\rho\beta})\Lambda_{\rho\sigma} \tag{38}%
\end{align}
Substituting eqns.(36) through (38) into eqn.(28), we get%

\begin{align*}
&  \partial_{\mu}\{\sqrt{-g(x)}[T_{\lambda}^{\mu}(\eta^{\lambda\rho}x^{\sigma
}-\eta^{\lambda\sigma}x^{\rho})]+\sqrt{-g(x)}[t_{\lambda}^{\mu}(\eta
^{\lambda\rho}x^{\sigma}-\eta^{\lambda\sigma}x^{\rho})]\\
&  +\sqrt{-g(x)}\frac{1}{16\pi G}[(\frac{\partial R}{\partial\partial_{\mu
}g^{\alpha\beta}(x)}-\partial_{\nu}\frac{\partial R}{\partial\partial_{\mu
}\partial_{\nu}g^{\alpha\beta}(x)}+\\
&  \frac{1}{2}g_{\xi\eta}\partial_{\nu}g^{\xi\eta}\frac{\partial R}%
{\partial\partial_{\mu}\partial_{\nu}g^{\alpha\beta}(x)})(\eta^{\beta\rho
}g^{\alpha\sigma}-\eta^{\beta\sigma}g^{\alpha\rho}+\eta^{\alpha\rho}%
g^{\sigma\beta}-\eta^{\alpha\sigma}g^{\rho\beta})\\
&  +\frac{\partial R}{\partial\partial_{\mu}\partial_{\nu}g^{\alpha\beta}%
(x)}(\eta^{\beta\rho}\partial_{\nu}g^{\alpha\sigma}-\eta^{\beta\sigma}%
\partial_{\nu}g^{\alpha\rho}+\eta^{\alpha\rho}\partial_{\nu}g^{\sigma\beta
}-\eta^{\alpha\sigma}\partial_{\nu}g^{\rho\beta})
\end{align*}%
\begin{equation}
-\frac{\partial R}{\partial\partial_{\mu}\partial_{\sigma}g^{\alpha\beta}%
(x)}\eta^{\lambda\rho}\partial_{\lambda}g^{\alpha\beta}+\frac{\partial
R}{\partial\partial_{\mu}\partial_{\rho}g^{\alpha\beta}(x)}\eta^{\lambda
\sigma}\partial_{\lambda}g^{\alpha\beta}]\}=0 \tag{39}%
\end{equation}
The expression in the first bracket is just the angular momentum tensor of the
matter fields. Hence the expression in the second bracket seems to be the
gravitational orbital angular momentun, and the expression in the third
bracket with its coefficient seems to be the gravitational spin angular momentun.

\subsection{Conservation of Electric Charge}

Action (7) for the complex scalar field in curved spacetime remains unchanged
under the following infinitesimal transformation of field.%

\begin{equation}
\delta x^{\mu}=0,\delta\varphi(x)=i\epsilon e\varphi(x),\delta\varphi^{\ast
}(x)=-i\epsilon e\varphi^{\ast}(x),\delta g^{\alpha\beta}(x)=0 \tag{40}%
\end{equation}
Substituting eqns.(13), (32) and (40) into eqn.(28), we get electric charge conservation,%

\begin{equation}
\partial_{\mu}[\sqrt{-g(x)}J^{\mu}(x)]=\sqrt{-g(x)}\nabla_{\mu}J^{\mu}(x)=0
\tag{41}%
\end{equation}
where%

\begin{equation}
J^{\mu}(x)=ieg^{\mu\nu}(x)[\varphi^{\ast}\nabla_{\nu}\varphi-(\nabla_{\nu
}\varphi^{\ast})\varphi] \tag{42}%
\end{equation}
is the $4-$vector of electric current density.

Now, we have obtained, by using Noether's theorem, the conservation laws of
`energy-momentum', `angular momentum', and electric charge. However, I am not
going to investigate the obtained energy-momentum, angular momentum densities
in detail here. Because I think only the last conservation law is
geometrically and physically meaningful. The reason will be given in next section.

\section{No-Go Theorem for Conservation of Energy-Momentum in Curved
Spacetime}

\subsection{Some preliminaries from geometry}

We will denote by $\mathcal{M}$ the spacetime manifold. For any spacetime
point (event) $p\in\mathcal{M}$, denote by $V_{p}$ the tangent space to
$\mathcal{M}$ at $p$, denote by $V_{p}^{\ast}$ the dual space of $V_{p}$ and
denote by $V_{p}^{\otimes r}$ the tensor product%

\begin{equation}
V_{p}\otimes V_{p}\otimes\cdot\cdot\cdot\otimes V_{p}\equiv V_{p}^{\otimes r}
\tag{43}%
\end{equation}
There are r copies of $V_{p}$ on the lhs.

Tangent vectors at different points ($p,q\ldots$) of a (generalized)
Riemannian manifold belong to different tangent spaces ($V_{p},V_{q}\ldots$).
One can not add up vectors belonging to different linear spaces. In order to
add them up, one has to parallelly transport them into one and the same
tangent space first. Parallel transport of vectors is coordinate free.
However, it depends on path, if the (generalized) Riemannian\ manifold is not
flat. Therefore, one cannot add up tangent vectors at different points of a
curved (generalized) Riemannian manifold in an objective way without any
subjective factor.

But in an affine space, which can be regarded as a flat (generalized)
Riemannian manifolds, parallel transport of tensors is coordinate free and
independent of path. Adding up tangent vectors at different points always
makes sense. Besides, in an affine coordinate system, the equation for
parallel transport of vector $v$ is trivial, $\delta v^{\gamma}=0$, hence the
$\nu-$component of sum vector is the sum of the $\nu-$components of addend
vectors, which remain unchanged during parallel transport. \textbf{In a
curvilinear coordinate system, however, the value of }$\nu-$\textbf{component
of sum vector is by no means the sum of }$\nu-$\textbf{components of addend
vectors before being parallelly transported. The latter is meaningless. }In
particular,\textbf{ }$\int_{\partial\Omega}ds_{\mu}\sqrt{-g(x)}T^{\mu\nu}(x)$,
the sum of $ds_{\mu}\sqrt{-g(x)}T^{\mu\nu}(x)$ ($\nu-$components of addend
vectors $dP$ before being parallelly transported) is meaningless.

Scalars at different points ($p,q\ldots$) of a (generalized) Riemannian
manifold belong to different linear spaces ($V_{p}^{\otimes0},V_{q}^{\otimes
0}\ldots$) too. Conceptually, before adding them up, one has to parallelly
transport them to one and the same point of the manifold. Parallel transport
of scalars is trivial ($\delta\phi=0$), coordinate free and path-independent.
Therefore, adding up scalars at different points of a (generalized) Riemannian
manifold is always meaningful, and the value of sum scalar is just the sum of
values of addend scalars, independent of where the sum scalar lies. In
particular, $\int_{\partial\Omega}ds_{\mu}\sqrt{-g(x)}J^{\mu}(x)$, the sum of
$ds_{\mu}\sqrt{-g(x)}J^{\mu}(x)$ (values of addend scalars $dQ$ ) is meaningful.

The above argument shows

\begin{proposition}
(No-Go Theorem) In a curved (generalized) Riemannian manifold, one cannot add
up the tangent vectors at different points of the manifold in a way free of
subjective options.
\end{proposition}

\begin{proposition}
(No-Go Theorem) In a curvilinear coordinate system, sum of $\nu-$components of
tangent vectors\ at different points of a (generalized) Riemannian manifold is
meaningless, no matter the manifold is flat or curved.
\end{proposition}

These propositions are well-known to geometers. Unfortunately, however,
neglect of them is the root of all the difficulties in energy-momentum
conservation in GR.

Now, let us get to the conservation laws.

\subsection{Inadequate Expression for Conservation of a Vector in Curved
Spacetime}

Let's first consider a tensor field $T$ of type $(2,0)$ with null covariant
divergence on Minkowski space. This tensor field $T$ can be considered the
density and current density of some vector, say, $P$ (not necessarily
energy-momentum 4-vector). In any Lorentzian coordinate system $(x^{0}%
,x^{1},x^{2},x^{3})$, we have%

\begin{equation}
\sqrt{-g(x)}\nabla_{\mu}T^{\mu\nu}(x)=\partial_{\mu}T^{\mu\nu}(x)=0 \tag{44}%
\end{equation}
Integrating (44) over 4-dimensional spacetime region $\Omega$, we get%

\begin{equation}
\int_{\Omega}d^{4}x\sqrt{-g(x)}\nabla_{\mu}T^{\mu\nu}(x)=\int_{\Omega}%
d^{4}x\partial_{\mu}T^{\mu\nu}(x)=\int_{\partial\Omega}ds_{\mu}T^{\mu\nu}(x)=0
\tag{45}%
\end{equation}
Choosing $\Omega$ such that $\partial\Omega$ consist of two pieces of
space-like hypersurface, $\Sigma_{2}$ on the top (future), and $\Sigma_{1}$ on
the bottom (past), and one piece of hypersurface $\Sigma$ with Lorentzian
signatured reduced metric in between, we have%

\begin{equation}
\int_{\Sigma_{2}}ds_{\mu}T^{\mu\nu}(x)-\int_{\Sigma_{1}}ds_{\mu}T^{\mu\nu
}(x)=-\int_{\Sigma}ds_{\mu}T^{\mu\nu}(x) \tag{46}%
\end{equation}
Eqn.(46) is considered in SR by all physicists as the law of conservation of
vector\textbf{ }$P$: The value of $P^{\nu}$ on space-like hypersurface
$\Sigma_{2}$ is equal to the value of $P^{\nu}$ on space-like hypersurface
$\Sigma_{1}$ plus the amount of $P^{\nu}$ which flows in through the boundary
during the corresponding evolution. Note that in this interpretation,
$\Sigma_{1}$ and $\Sigma_{2}$\ are not necessarily hyperplanes, while
coordinate system $(x^{0},x^{1},x^{2},x^{3})$ must be Lorentzian.

Now let us switch to curvilinear coordinate system $(y^{0},y^{1},y^{2},y^{3}%
)$, while keeping the tensor field and spacetime unchanged (the spacetime is
still Minkowski space and the covariant divergence of $T$ is still null).
Eqn.(44) becomes%

\begin{align}
\sqrt{-g(y)}\nabla_{\mu}T^{\mu\nu}(y)  &  =\sqrt{-g(y)}[\partial_{\mu}%
T^{\mu\nu}(y)+\Gamma_{\mu\lambda}^{\mu}(y)T^{\lambda\nu}(y)+\Gamma_{\mu
\lambda}^{\nu}(y)T^{\mu\lambda}(y)]\nonumber\\
&  =\partial_{\mu}[\sqrt{-g(y)}\partial_{\mu}T^{\mu\nu}(y)]+\sqrt{-g(y)}%
\Gamma_{\mu\lambda}^{\nu}(y)T^{\mu\lambda}(y)=0 \tag{47}%
\end{align}
Integrating\ $\partial_{\mu}[\sqrt{-g(y)}\partial_{\mu}T^{\mu\nu}(y)]$ over
4-dimensional spacetime region $\Omega$, we get\ \ \ \ %

\begin{align}
\int_{\Omega}d^{4}y\partial_{\mu}[\sqrt{-g(y)}T^{\mu\nu}(y)]  &  =\int
_{\Sigma_{2}}ds_{\mu}\sqrt{-g(y)}T^{\mu\nu}(y)-\int_{\Sigma_{1}}ds_{\mu}%
\sqrt{-g(y)}T^{\mu\nu}(y)\nonumber\\
+\int_{\Sigma}ds_{\mu}\sqrt{-g(y)}T^{\mu\nu}(y)  &  =-\int_{\Omega}%
d^{4}y[\sqrt{-g(y)}\Gamma_{\mu\lambda}^{\nu}(y)T^{\mu\lambda}(y)]\neq0
\tag{48}%
\end{align}
Einstein and his followers on this subject consider $\int_{\partial\Omega
}ds_{\mu}\sqrt{-g(y)}T^{\mu\nu}(y)=0$, or equivalently $\partial_{\mu}%
(\sqrt{-g(y)}T^{\mu\nu}(y))=0$ as the integral or differential conservation
law of vector $P$. That is, they consider $\int_{\Sigma_{1}}ds_{\mu}%
\sqrt{-g(y)}T^{\mu\nu}(y)$ as the total value of $P^{\nu}$ over space-like
hypersurface $\Sigma_{1}$. But, as has been shown above, this is right only in
affine coordinate systems, and there is no such coordinate system in curved
spacetime. In curvilinear coordinate systems, the $\nu-$component of the sum
of vectors $dP$'s over space-like hypersurface $\Sigma_{1}$ differs from the
sum of $(dP)^{\nu}=ds_{\mu}\sqrt{-g(y)}T^{\mu\nu}(y)$ over space-like
hypersurface $\Sigma_{1}$. The former is geometrically meaningful, the latter
is meaningless. $\int_{\Omega}d^{4}y\partial_{\mu}[\sqrt{-g(y)}T^{\mu\nu
}(y)]=0$\textbf{ is an inadequate expression for conservation of vector }%
$P$\textbf{.} It was just this misunderstanding that has guided them in
searching for gravitational energy-momentum tensor to keep the total
energy-momentum conserved. According to their point of view, inequality (48)
would be read as vector $P$ is not conserved even in Minkowski space. Thus,
originally conserved vector $P$ in Minkowski space is no longer conserved
without any geometrical or physical change, just due to switching to
curvilinear coordinate system. This is obviously unreasonable, since the
objective law of nature should be coordinate free,\ and a\ proper
generalization of a proposition in SR to GR should retrieve it in the flatness limit.\ 

\subsection{Conclusion}

Einstein and some others realized the difficulty of energy-momentum
conservation in GR. They take the law of conservation of energy-momentum for
granted, consider $\int_{\Omega}d^{4}y\partial_{\mu}[\sqrt{-g(y)}T^{\mu\nu
}(y)]=0$\textbf{ }as the adequate expression for conservation of
energy-momentum for matter field. In order to keep the total energy-momentum
conservation alive, they introduce the energy-momentum psuedotensor of
gravity. After ninety years of efforts, people realized there is no local
solution to this question.

According to the no-go theorems proposed above, I think it is impossible to
define the total energy over a piece of space-like hypersurface, hence it is
meaningless to talk about energy-momentum conservation in curved spacetime.
Gravitational field ( I'd rather call it metric field) is different to matter
fields. It shouldn't carry energy.

The correctness of any proposition in natural science has to be tested by
experiments. Up to now, no evidence for gravity carrying energy has been
tested experimentally yet. The observed accelerated expansion of universe
seems to support non-conservation of energy-momentum.

The law of conservation of energy-momentum still works rather well in regions
of spacetime which is approximately flat, like the part of universe nearby.

\begin{acknowledgement}
This work has been supported by (Zhao Zhan-)Yue-(Pan Ye-)Hong Science
foundation. I thank Prof. Zhan-Yue Zhao, Prof. Shi-Hao Chen and Prof. Guo-Ying
Qi for helpful discussions.
\end{acknowledgement}


\begin{thebibliography}{99}                                                                                               %


\bibitem {1}A. Einstein, \textit{Sber. preuss}. 788(1915); 448(1918).

\bibitem {2}H. Bauer, \textit{Phys. Z.} \textbf{19}, 163(1918).

\bibitem {3}R.C. Tolman, \textit{Phys. Rev.} \textbf{35, }875(1930).

\bibitem {4}L.D. Landau and E.M. Lifshitz, \textit{The classical theory of
fields, 4th ed., }(Butterworth-Heinemann, Beijing, 1999) \textit{P.}280.

\bibitem {5}A. Papapetrou, Proc. R. Irish Acad. \textbf{A52, }11(1948).

\bibitem {6}C. M\"{o}ller, \textit{Annals of Physics} \textbf{4, }347(1958);
\textit{Annals of Physics} \textbf{12, }118(1961).

\bibitem {7}S. Weinberg, \textit{Gravitation and Cosmology }(Wiley, New York,
1972), P.165.

\bibitem {8}P.G. Bergmann and R. Thompson, \textit{Phys. Rev.} \textbf{D89},
400 (1953).

\bibitem {9}H. Bondi, \textit{Proc. R. Soc. London }\textbf{A427}, 249(1990).

\bibitem {10}R. Penrose,\textit{Proc. R. Soc. London }\textbf{A388}, 457(1982).

\bibitem {11}M.B. Mensky, \textit{Phys. Lett. }\textbf{A328}, 261(2004).
\end{thebibliography}
\end{document}